\documentstyle[preprint,aps]{revtex}
\tightenlines
\begin{document}
\draft
\title
{Hausdorff dimension and anyonic distribution functions}
\author
{Wellington da Cruz\footnote{E-mail: wdacruz@fisica.uel.br}} 
\address
{Departamento de F\'{\i}sica,\\
 Universidade Estadual de Londrina, Caixa Postal 6001,\\
Cep 86051-970 Londrina, PR, Brazil\\}
\date{\today}
\maketitle
\begin{abstract}
We obtain the distribution functions for anyonic excitations classified 
into equivalence 
classes labeled by Hausdorff dimension, $h$ and as an example 
of such anyonic systems, we consider the 
collective excitations of the Fractional Quantum Hall Effect ( FQHE ).  
\end{abstract}



We have obtained in\cite{R1} from a continuous family of 
Lagrangians for fractional spin particles a path integral 
representation for the propagator and its representation in 
moment space. On the other hand, the trajectories swept out 
by scalar and spinning particles can be characterized by 
the fractal parameter $h$ ( or the Hausdorff dimension ). We have 
that, $L$, 
the length of closed trajectory with size $R$ has its fractal 
properties described by $L\sim R^h$\cite{R2}, such that 
for scalar particle 
 $L\sim \frac{1}{p^2}$, $R^2\sim L$ and $h=2$ ; for  
 spinning particle, $
 L\sim \frac{1}{p}$, $R^1\sim L$ and $h=1$. From the form of 
 anyonic propagator given in\cite{R1}, with the spin defined 
 into the interval $0\leq s\leq 0.5$, we have extracted the 
 following formula, $h=2-2s$, which relates the Hausdorff 
 dimension $h$ and the spin $s$ of the particle. Thus, 
 for anyonic particle, $h$ takes values within the interval 
 $1$$\;$$ < $$\;$$h$$\;$$ <$$\;$$ 2$. 
 In \cite{R3}, we have classified 
 the fractional 
spin particles or anyonic excitations in terms of 
equivalence classes labeled by $h$. 
Therefore, such particles in a specific class 
can be considered on equal footing.

On the other hand, in the context of FQHE systems, the 
filling factor, 
a parameter which characterize that phenomenon, 
can be classified in this 
terms too. We have, therefore, a new hierarchy scheme 
for the filling factors\cite{R4}, which is 
extracted from the relation between $h$ and the 
statistics $\nu$ 
( or the filling factors ). Our approach, contrary to 
literature\cite{R5} do not have an empirical character 
and we 
can predicting for which values of $\nu$ FQHE can be 
observed. 
A Braid group structure behind this classification 
also was 
noted\cite{R6} and in\cite{R3} a topological invariant, 
${\cal W}=h+2s-2p$, was introduced which relates 
a characteristic 
of the multiply connected spaces with the numbers of 
fractional spin particles, $p$, and the quantities 
related 
to the particles, $h$ and $s$ ( note that this invariant
 has a connection with the {\it Euler characteristic} of the 
 surface ). The anyonic model, a 
charge-flux 
system, constitutes a topological obstruction 
( holes with hard core repulsion ) in this description and 
the elements of the Braid group are 
equivalents trajectories. 
 
 Now, we propose a statistical weight for 
such excitations in terms 
of $h$, as follows:
  
\begin{equation}
\label{e.1}
\omega_{j}=\frac{\left[G_{j}+(N_{j}-1)(h-1)\right]!}{N_{j}!
\left[G_{j}+(N_{j}-1)(h-1)-N_{j}\right]!},
\end{equation}

\noindent where $G_{j}$ means a group of quantum states, 
$N_{j}$ is the number 
of particles and for, $h=2$ we have bosons and for 
$h=1$ we have fermions. For, 
$1$$\;$$ < $$\;$$h$$\;$$ <$$\;$$ 2$, we have anyonic 
excitations which interpolates between these two extremes. 
The fractal parameter $h_{i}$ 
 ( $i$ means a specific interval ) is related 
 to statistcs $\nu$, in the 
  following way:  

\begin{eqnarray}
\label{e.2}
&&h_{1}=2-\nu,\;\;\;\; 0 < \nu < 1;\;\;\;\;\;\;\;\;
 h_{2}=\nu,\;\;\;\;
\;\;\;\;\;\;\;\;\; 1 <\nu < 2;\;\nonumber\\
&&h_{3}=4-\nu,\;\;\;\; 2 < \nu < 3;\;\;\;\;\;\;\;\;
h_{4}=\nu-2,\;\;\;\;\;\;\; 3 < \nu < 4;\;\nonumber\\
&&h_{5}=6-\nu,\;\;\;\; 4 < \nu < 5;\;\;\;\;\;\;\;
h_{6}=\nu-4,\;\;\;\;\;\;\;\; 5 < \nu < 6;\;\\
&&h_{7}=8-\nu,\;\;\;\; 6 < \nu < 7;\;\;\;\;\;\;\;
h_{8}=\nu-6,\;\;\;\;\;\;\;\; 7 < \nu < 8;\;\nonumber\\
&&h_{9}=10-\nu,\;\;8 < \nu < 9;\;\;\;\;\;\;
h_{10}=\nu-8,\;\;\;\;\;\; 9 < \nu < 10;\nonumber\\
&&etc,\nonumber
\end{eqnarray}

\noindent such that this spectrum of $\nu$, as we can see, 
has a 
complete mirror symmetry and for a given $h$, we collect
 different values of $\nu$ in a specific class, for example, 
 consider $h=\frac{5}{3}$ and $h=\frac{4}{3}$, so we obtain;
 $\left\{\frac{1}{3},\frac{5}{3},\frac{7}{3},
 \frac{11}{3},\frac{13}{3},\frac{17}{3},\cdots\right\}_
{h=\frac{5}{3}}$ and  $\left\{\frac{2}{3},
\frac{4}{3},\frac{8}{3},\frac{10}{3},\frac{14}{3},
\frac{16}{3},
\cdots\right\}_
{h=\frac{4}{3}}$.

These particles or excitations, in the class, share some 
common 
characteristics, according our interpretation, in the 
same way that bosons and fermions. We observe that 
each interval 
contributes with one and only one particle to the 
class and the 
statistics and the spin, are related by $\nu=2s$. 

We stress that, our expression for statistical 
weight 
Eq.(\ref{e.1}), is more general than an expression given 
in\cite{R7},
once if we consider the relation between $h$ and $\nu$, we 
just obtain that first result and extend it for the 
complete spectrum of statistics $\nu$, for example: 
 
\begin{eqnarray}
\label{e.4}
&&\omega_{j}=\frac{\left[G_{j}+(N_{j}-1)(1-\nu)\right]!}
{N_{j}!\left[G_{j}+(N_{j}-1)(1-\nu)-N_{j}\right]!},
 \;\;\;\; 0 < \nu < 1; \nonumber\\
&&\omega_{j}=\frac{\left[G_{j}+(N_{j}-1)(\nu-1)\right]!}
{N_{j}!\left[G_{j}+(N_{j}-1)(\nu-1)-N_{j}\right]!},
\;\;\;\; 1 < \nu < 2; \\
&&\omega_{j}=\frac{\left[G_{j}+(N_{j}-1)(3-\nu)\right]!}
{N_{j}!\left[G_{j}+(N_{j}-1)(3-\nu)-N_{j}\right]!},
\;\;\;\; 2 < \nu < 3;\nonumber\\
&&\omega_{j}=\frac{\left[G_{j}+(N_{j}-1)(\nu-3)\right]!}
{N_{j}!\left[G_{j}+(N_{j}-1)(\nu-3)-N_{j}\right]!},
\;\;\;\; 3 < \nu < 4;\nonumber\\
&&etc.\nonumber
\end{eqnarray}

\noindent  The expressions 
Eq.(\ref{e.4}) were possible because of the mirror symmetry 
as just have said above. But, our approach 
in terms of Hausdorff dimension, $h$, have an advantage, 
because we collect into equivalence class the anyonic 
excitations and so, we consider on equal footing the 
excitations 
in the class and this is a new approach which enter on 
considerations about fractional spin particles. For different 
species of particles, the statistical weight take the form

\begin{equation}
\label{e.5}
\Gamma=\prod_{j}\omega_{j}=\prod_{j}\frac{\left
[G_{j}+(N_{j}-1)
(h-1)\right]!}{N_{j}!\left[G_{j}+(N_{j}-1)(h-1)-N_{j}\right]!},
\end{equation}
\noindent which reduces to Eq.(\ref{e.1}) for 
only one species. Now, we can consider the entropy 
for a given class $h$. Taking the logarithm of 
Eq.(\ref{e.5}), 
with the condition that, $N_{j}$ and $G_{j}$ are vary 
large numbers and 
defining the average occupation numbers, $n_{j}=
\frac{N_{j}}{G_{j}}$, 
we obtain for a gas not in equilibrium, an expression 
for the entropy

\begin{eqnarray}
&&{\cal S}=\sum_{j}G_{j}\left\{\left[1+n_{j}(h-1)
\right]\ln\left(
\frac{1+n_{j}(h-1)}{1+n_{j}(h-1)-n_{j}}\right)\right.\\
&&-\left.n_{j}\ln\left(\frac{n_{j}}{1+n_{j}(h-1)-n_{j}}
\right)\right\},
\nonumber
\end{eqnarray}

\noindent such that for $h=2$ and $h=1$, we obtain
the expressions for a Bose and Fermi 
gases not in equilibrium\cite{R8}, 
  
\begin{eqnarray}
&&{\cal S}=\sum_{j}G_{j}\left\{\left(1+n_{j}\right)\ln
\left(1+n_{j}\right)-n_{j}\ln n_{j}\right\};\\
&&{\cal S}=-\sum_{j}G_{j}\left\{n_{j}\ln n_{j}+
\left(1-n_{j}\right)\ln \left(1-n_{j}\right)\right\},
\end{eqnarray}
\noindent respectively.

The distribution function for a gas of the class $h$ can 
be obtained from the condition of the entropy be a 
maximum. Thus, 
we have

\begin{eqnarray}
\label{e.9}
n_{j}\xi=\left\{1+n_{j}(h-1)\right\}^{h-1}\left\{1+n_{j}
(h-2)\right\}^{2-h},
\end{eqnarray}

\noindent where $\xi=\exp\left\{(\epsilon_{j}-\mu)/KT\right\}$, 
has the usual definition and the Bose and Fermi distributions
 are obtained for values of $h=2,1$; respectively.

In the 
context of the 
Fractional Quantum Hall Effect ( FQHE ), as we said 
in the introduction, 
the filling factor ( rational number with an 
odd denominator ), 
can be also classified in 
terms of $h$\cite{R4}. We have that the anyonic excitations 
are collective 
excitations manifested as quasiparticles or quasiholes in FQHE 
systems. 
Thus, for example, we have the collective 
excitations as given in 
the beginning for $h=\frac{5}{3}$ and 
$h=\frac{4}{3}$. Now, we 
have noted 
that these collections contain filling fractions, in 
particular, $\nu=
\frac{1}{3}$ and $\nu=\frac{2}{3}$, that experimentally 
were observed\cite{R9} and so we are 
able to estimate for which values of $\nu$ the largest 
charge gaps occurs, or alternatively, we can {\it predicting} 
FQHE. As was 
observed in\cite{R4} this is a new schem hierarchy scheme for 
the filling 
factors, that express the occurrence of the FQHE in more 
fundamental terms, that is, relating the fractal parameter $h$ 
and the filling factors $\nu$. Therefore, for 
anyonic excitations in a stronger 
magnetic field 
at low temperatures, our generalization 
( Eq.\ref{e.4} ) of the results 
in\cite{R7} can be 
considered. 

In summary, we have obtained distribution 
functions ( Eq.\ref{e.9} ) for 
anyonic excitations in terms of the Hausdorff dimension 
$h$, which classified the anyonic excitations into 
equivalence classes and reduces to 
fermionic and bosonic distributions, when $h=1$ and 
$h=2$, respectively. In this way, we have extended
 ( Eq.\ref{e.4} ) results 
of the literature\cite{R7} 
for the complete spectrum of 
statistics $\nu$. A connection with the FQHE, 
considering the filling factors into equivalence classes 
labeled by $h$ also was 
considered and an estimate for occurrence of FQHE made.


\begin{thebibliography}{99}
\bibitem{R1} W. da Cruz, preprint/UEL-DF/W-01/97, hep-th/9803020.
\bibitem{R2} A. M. Polyakov, in {\it Proc. Les
 Houches Summer School
 {\bf vol. IL}}, ed. E. Br\'ezin and J. Zinn-Justin
  (North Holland, 1990), 305. 
\bibitem{R3} W. da Cruz, preprint/UEL-DF/W-02/97, hep-th/9803043; 
ibid, preprint/UEL-DF/W-03/97, hep-th/9802135.
\bibitem{R4} W. da Cruz, preprint/UEL-DF/W-04/97, cond-mat/9802266.
\bibitem{R5} F. D. M. Haldane, Phys. Rev. Lett. 
{\bf 51} (1983) 605; B. I. Halperin, Phys. Rev. Lett. 
{\bf 52} (1984) 1583; see also\cite{R9}.
\bibitem{R6} W. da Cruz, preprint/UEL-DF/W-05/97, cond-mat/9802267.  
\bibitem{R7} Y. S. Wu, Phys. Rev. Lett. {\bf 73} 
(1994) 922 .
\bibitem{R8} E. M. Lifshitz and L. P. Pitaevskii, 
{\it Statistical Physics, Part 1, 
3rd. edition } ( Pergamon Press, Oxford, 1980 ).
\bibitem{R9} A. H. MacDonald, cond-mat/9410047; 
C. Cros and A. H. MacDonald, 
Phys. Rev. {\bf B 42} (1990) 10811 and 
references therein.  
\end{thebibliography}
\end{document}